\documentclass[a4paper]{jpconf}
\usepackage{graphicx}
\usepackage{amsmath,amssymb}
\newcommand{\bra}[1]{\langle {#1} |}
\newcommand{\ket}[1]{| {#1} \rangle}

\begin{document}
\title{Finite amplitude method in linear response TDDFT calculations}

\author{Takashi Nakatsukasa}

\address{RIKEN Nishina Center, Wako 351-0198, Japan}
\address{Center for Computational Sciences, University of Tsukuba,
Tsukuba, 305-8571, Japan}

\ead{nakatsukasa@riken.jp}

\begin{abstract}
The finite amplitude method is a feasible and efficient method for
the linear response calculation based on the time-dependent
density functional theory.
It was originally proposed as a method to calculate the strength functions.
Recently, new techniques have been
developed for computation of normal modes (eigenmodes) of the
quasiparticle-random-phase approximation.
Recent advances associated with the finite amplitude method are reviewed.
\end{abstract}

\section{Introduction}

The nuclear energy density functional methods are extensively utilized
in studies of nuclear structure and reaction \cite{BHR03}.
For the linear response calculations,
the quasiparticle-random-phase
approximation (QRPA), based on the time-dependent
density-functional theory \cite{NY01,NY05,Nak12,Sim12},
is a standard theory \cite{RS80}.
However, for realistic energy functionals,
it demands both complicated coding and large-scale computational
resources.
To resolve these issues,
there have been several developments \cite{JBH99,Toi10,CTP12}, including
the finite amplitude method (FAM) \cite{NIY07}.

The FAM allows us to avoid explicit evaluation of
complex residual fields, which significantly reduces the necessary
coding effort for realistic energy functionals.
In addition, the use of iterative solution of the FAM equations
also reduces the computational task and the memory requirement.
The FAM was first adopted for the linear response calculations for
the electric dipole mode,
using the Skyrme energy functionals without the pairing correlations
\cite{INY09,INY11}.
The method was soon extended to superfluid systems \cite{AN11}.
Then, the FAM was adopted in the code {\sc hfbtho} to study monopole
resonances in superfluid deformed nuclei with axially symmetry.
Mario Stoitsov played a leading role in this project,
which started during his visit to RIKEN in 2010 \cite{Sto11}.
Later, there have been further developments in the FAM with
relativistic and non-relativistic frameworks \cite{AN13,LNZM13,HKN13}

The FAM was originally proposed for a feasible method to calculate
the response function of a given one-body operator.
Recently, new methodologies have been developed, for calculating
discrete eigenstates using the FAM.
In this article, we review some of these recent developments.

\section{The finite amplitude method}

In this section, we briefly illustrate the essential idea of the FAM.
Let us start from the Hartree-Fock-Bogoliubov (HFB) equations,
\begin{equation}
H \Phi_i = E_i \Phi_i , \quad\quad
H \tilde{\Phi}_i = -E_i \tilde{\Phi}_i ,
\end{equation}
where $H$, $E_i>0$, and $\Phi_i=\begin{pmatrix} U_i \\ V_i\end{pmatrix}$ are
the HFB Hamiltonian including the particle-number cranking term ($-\mu N$),
the quasiparticle energies, and
the quasiparticle states associated with the HFB ground state, respectively.
$\tilde{\Phi}_i=\begin{pmatrix} V_i^* \\ U_i^*\end{pmatrix}$ are 
conjugate to $\Phi_i$ with negative quasiparticle energies \cite{RS80}.
They are orthonormalized as
$\Phi_i^\dagger\Phi_j=\tilde\Phi_i^\dagger \tilde\Phi_j=\delta_{ij}$
and
$\Phi_i^\dagger\tilde\Phi_j=\tilde\Phi_i^\dagger \Phi_j=0$.
The generalized density matrix $R$ \cite{RS80} at the ground state
can be written in a simple form as
\begin{equation}
R\equiv
\begin{pmatrix}
\rho & \kappa \\
-\kappa^* & 1-\rho^*
\end{pmatrix}
= 1 - \sum_i \Phi_i \Phi_i^\dagger
= \sum_i \tilde{\Phi}_i \tilde{\Phi}_i^\dagger .
\end{equation}
In the small-amplitude (linear) approximation,
the density fluctuation $\delta R$ can be
decomposed into normal modes $\delta R^{(n)}$,
\begin{equation}
\delta R(t)= \sum_n \delta R^{(n)} e^{-i\omega_n t} + \textrm{h.c.} .
\end{equation}
In the linear order, the matrix elements
of $\delta R^{(n)}$ between $\Phi_i$ and $\tilde\Phi_j$ are denoted by
$X_{ij}^{(n)}$ and $Y_{ij}^{(n)}$ as
\begin{eqnarray}
&&X_{ij}^{(n)} \equiv \Phi_i^\dagger \delta R^{(n)} \tilde{\Phi}_j
= - \Phi_j^\dagger \delta R^{(n)} \tilde{\Phi}_i
= -X_{ji}^{(n)}, \\
&&Y_{ij}^{(n)} \equiv \tilde\Phi_j^\dagger \delta R^{(n)} \Phi_i
= -\tilde\Phi_i^\dagger \delta R^{(n)} \Phi_j
=- Y_{ji}^{(n)}
.
\end{eqnarray}
The other matrix elements of $\delta R$ vanish
in the quasiparticle basis:
\begin{equation}
\Phi_i^\dagger \delta R^{(n)} \Phi_j=
\tilde\Phi_i^\dagger \delta R^{(n)} \tilde\Phi_j =0 .
\end{equation}
Note that $\delta R^{(n)}$ are in general non-Hermitian, nevertheless
$\delta R(t)$ is Hermitian.

The density fluctuation $\delta R^{(n)}$ induces an residual field in the
HFB Hamiltonian, $H=H+\delta H^{(n)}$,
where $\delta H^{(n)}$ linearly depends on $\delta R^{(n)}$.
The essential idea of the FAM is that, instead of explicitly expanding
$\delta H^{(n)}$ in the linear order in $\delta R^{(n)}$,
we compute $\delta H^{(n)}$ by the finite difference
using a small parameter $\eta$ as
\begin{equation}
\label{delta_H}
\delta H^{(n)}=\frac{1}{\eta}\left(
H[R+\eta\delta R^{(n)}]-H 
\right) .
\end{equation}
The Hamiltonian $H[R+\eta \delta R^{(n)}]$ should be evaluated at
the density,
$
R+\eta\delta R^{(n)}= 1-\sum_i \Psi'_i \Psi_i^\dagger
 = \sum_i \tilde\Psi'_i \tilde\Psi_i^\dagger,
$
using the following quasiparticles
\begin{equation}
\label{FAM_qp_1}
\Psi_i^\dagger\equiv\Phi_i^\dagger
-\eta \sum_j X_{ij}^{(n)} \tilde\Phi_j^\dagger ,
\quad\quad
\Psi'_i\equiv\Phi_i-\eta \sum_j Y_{ij}^{(n)} \tilde\Phi_j ,
\end{equation}
or
\begin{equation}
\label{FAM_qp_2}
\tilde\Psi_i^\dagger\equiv \tilde\Phi_i^\dagger
+\eta \sum_j \Phi_j^\dagger Y_{ji}^{(n)} ,
\quad\quad
\tilde\Psi'_i\equiv \tilde\Phi_i+\eta \sum_j \Phi_j X_{ji}^{(n)} .
\end{equation}
For a given set of forward and backward amplitudes, $\{X^{(n)},Y^{(n)}\}$,
the calculation of Eq. (\ref{delta_H}) is relatively easy
because all we need to calculate are the one-body quantities with
two-quasiparticle indices, such as $\delta H_{ij}^{(n)}=\Phi_i^\dagger \delta H\tilde\Phi_j$.
In contrast, the QRPA matrices have four-quasiparticle indices \cite{RS80},
$A_{ij,kl}$ and $B_{ij,kl}$.
The calculation of these matrix elements demands significant coding effort.

\section{Calculation of the QRPA matrices}

In this section, we recapitulate the matrix-FAM (m-FAM) proposed in
Ref.~\cite{AN13}.
This provides a simple numerical method
to calculate the QRPA matrices
using the principle of the FAM, Eqs. (\ref{delta_H}), (\ref{FAM_qp_1}),
and (\ref{FAM_qp_2}).

The QRPA equation in the matrix form is given by
${\cal H} \vec{Z}_n =\omega_n {\cal N} \vec{Z}_n$, where
\begin{equation}
{\cal H}\equiv
\begin{pmatrix}
A & B \\
B^* & A^*
\end{pmatrix}
,\quad
{\cal H}\equiv
\begin{pmatrix}
1 & 0 \\
0 & -1
\end{pmatrix}
,\quad
\vec{Z}_n\equiv
\begin{pmatrix}
X^{(n)} \\
Y^{(n)}
\end{pmatrix}
.
\end{equation}
The most demanding part is the calculation of the matrix elements,
$A_{ij,kl}$ and $B_{ij,kl}$, which are formally defined by
the derivative of the HFB Hamiltonian with respect to the density.
However, since the FAM allows us to evaluate ${\cal H}\vec{Z}$
for a given vector $\vec{Z}$,
these matrix elements are provided by the FAM
in the following way.

Let us define the ``forward'' unit vector $\hat{e}_{kl}$
as $X_{ij}=\delta_{ik}\delta_{jl}$ and $Y_{ij}=0$,
and the ``backward'' one $\tilde{e}_{kl}$ as
$Y_{ij}=\delta_{ik}\delta_{jl}$ and $X_{ij}=0$.
Namely, these unit vectors are
\begin{equation}
\hat{e}_{kl}=
\begin{pmatrix}
X\\
Y
\end{pmatrix}
=
\begin{pmatrix}
0\\
\vdots \\
1 \\
0\\
\vdots
\end{pmatrix}
\begin{matrix}
\\
\leftarrow X_{kl}\\
\\
\end{matrix}
,\quad\quad
\tilde{e}_{kl}=
\begin{pmatrix}
X\\
Y
\end{pmatrix}
=
\begin{pmatrix}
0\\
\vdots \\
1 \\
0\\
\vdots
\end{pmatrix}
\begin{matrix}
\\
\leftarrow Y_{kl}\\
\\
\end{matrix}
.
\end{equation}
Then, it is trivial to see that
the upper component of $({\cal H}\vec{e}_n)_{ij}$
and 
$({\cal H}\tilde{e}_n)_{ij}$
are identical to
$A_{ij,kl}$ and $B_{ij,kl}$, respectively.
\begin{equation}
\left( {\cal H} \hat{e}_{kl} \right)_{ij}^\textrm{up}
= A_{ij,kl}
,
\quad\quad
\left( {\cal H} \tilde{e}_{kl} \right)_{ij}^\textrm{up}
= B_{ij,kl}
\end{equation}
On the other hand, using the FAM,
$({\cal H}\hat{e}_{kl})^\textrm{up}$ reads
\begin{equation}
A_{ij,kl}=
\left({\cal H}\hat{e}_{kl}\right)^\textrm{up}_{ij} = 
(E_i+E_j) \delta_{ik}\delta_{jl} + {\Phi_i^\dagger\delta H\tilde\Phi_j} 
,
\end{equation}
where $\Phi_i^\dagger\delta H\tilde\Phi_j$ in the right hand side
can be computed according to Eq. (\ref{delta_H}).
Here, $R+\eta\delta R$ is given by the ground-state
quasiparticles, $\Psi_i^\dagger=\Phi_i^\dagger$ and $\Psi'_i=\Phi_i$, 
except for $i=k$ and $l$,
\begin{equation}
\label{FAM_qp_3}
\Psi_k^\dagger=\Phi_k^\dagger
-\eta \tilde\Phi_l^\dagger ,
\quad
\Psi_l^\dagger=\Phi_l^\dagger
+\eta \tilde\Phi_k^\dagger .
\end{equation}
Following the same procedure with the ``backward'' $\tilde{e}_{kl}$,
we can calculate $B_{ij,kl}$.

The numerical coding of the present method is extremely easy.
All we need to calculate is the HFB Hamiltonian at the density
$R+\eta\delta R$ which is defined by Eq. (\ref{FAM_qp_3}).
After constructing the QRPA matrix, the QRPA normal modes of excitation
are obtained by diagonalizing the QRPA matrix \cite{RS80}.

\section{Iterative FAM with a contour integral in the complex frequency plane}

In this section, we present a contour integral technique combined with
the FAM developed in Ref.~\cite{HKN13}.

The iterative FAM (i-FAM) with a complex frequency $\omega$
provides a solution of the linear response equation
$( {\cal H} - \omega {\cal N} ) \vec{Z}(\omega) = -{\cal F}$,
where ${\cal F}$ is the one-body external field \cite{NIY07,AN11}.
\begin{equation}
{\cal F}\equiv
\begin{pmatrix}
F \\
F^*
\end{pmatrix}
,\quad
\hat{F}=\sum_{i<j} F_{ij} a_i^\dagger a_j^\dagger  + \textrm{h.c.} + \cdots
\end{equation}
Here, we assume $\hat{F}$ is a Hermitian operator.
The QRPA response function is calculated as \cite{NIY07}
\begin{equation}
S(F;\omega)=\sum_{i<j} \left\{
F_{ij}^* X_{ij}(\omega) +F_{ij} Y_{ij}(\omega)
\right\} .
\end{equation}
This can be expressed in terms of the QRPA normal modes as
\begin{equation}
S(F;\omega)=
\sum_n |\bra{n} \hat{F} \ket{0}|^2
\left(
\frac{1}{\omega-\omega_n} -\frac{1}{\omega+\omega_n}
\right) ,
\end{equation}
This shows that the transition strength for the $n$-th normal mode,
$|\bra{n}\hat{F}\ket{0}|^2$,
is a residue at $\omega=\omega_n$.
Therefore, if we choose the contour $C_n$ in the complex $\omega$-plane
that encloses $\omega=\omega_n>0$,
it reads
\begin{equation}
|\bra{n} \hat{F} \ket{0}|^2
= \frac{1}{2\pi i}\oint_{C_n} S(F;\omega) d\omega .
\end{equation}
The corresponding QRPA normal modes,
$X^{(n)}$ and $Y^{(n)}$,
are also given by the contour integral as \cite{HKN13}
\begin{eqnarray}
X^{(n)}_{ij}=
|\bra{n} \hat{F} \ket{0}|^{-1}
\frac{1}{2\pi i}\oint_{C_n} X_{ij}(\omega) d\omega ,\\
Y^{(n)}_{ij}=
|\bra{n} \hat{F} \ket{0}|^{-1}
\frac{1}{2\pi i}\oint_{C_n} Y_{ij}(\omega) d\omega .
\end{eqnarray}

The contour integral method with i-FAM is complementary to the m-FAM.
In the present approach, the eigenmodes are obtained by solving the
i-FAM equations for complex frequencies combined with the
contour integral.
In the m-FAM, we do not resort to an iterative algorithm to solve
the FAM equations, but we need to diagonalize the QRPA matrix at the end.
For a small model space, the m-FAM has a significant advantage, however,
the computational task of the m-FAM strongly
depends on the size of the QRPA matrix $D$,
which typically scales as $D^3$.


\ack
The present work is supported by JSPS KAKENHI Grant numbers
24105006, 25287065, and 25287066.
We thank all the collaborators on the present subject,
including Dr Mario Stoitsov.
We shall always remember him for his excellent contributions to
the field of theoretical and computational nuclear physics.

\section*{References}
\bibliographystyle{iopart-num}
\bibliography{nuclear_physics,myself}

\providecommand{\newblock}{}
\begin{thebibliography}{10}
\expandafter\ifx\csname url\endcsname\relax
  \def\url#1{{\tt #1}}\fi
\expandafter\ifx\csname urlprefix\endcsname\relax\def\urlprefix{URL }\fi
\providecommand{\eprint}[2][]{\url{#2}}

\bibitem{BHR03}
Bender M, Heenen P~H and Reinhard P~G 2003 {\em Rev. Mod. Phys.\/} {\bf 75}
  121

\bibitem{NY01}
Nakatsukasa T and Yabana K 2001 {\em J. Chem. Phys.\/} {\bf 114} 2550

\bibitem{NY05}
Nakatsukasa T and Yabana K 2005 {\em Phys. Rev. C\/} {\bf 71} 024301

\bibitem{Nak12}
Nakatsukasa T 2012 {\em Progress of Theoretical and Experimental Physics\/}
  {\bf 2012} 01A207

\bibitem{Sim12}
Simenel C 2012 {\em The European Physical Journal A\/} {\bf 48} 1

\bibitem{RS80}
Ring P and Schuck P 1980 {\em The nuclear many-body problems\/}
 (New York: Springer-Verlag)

\bibitem{JBH99}
Johnson C, Bertsch G and Hazelton W 1999 {\em Comp. Phys.
  Comm.\/} {\bf 120} 155

\bibitem{Toi10}
Toivanen J, Carlsson B~G, Dobaczewski J, Mizuyama K, Rodr\'iguez-Guzm\'an R~R,
  Toivanen P and Vesel\'y P 2010 {\em Phys. Rev. C\/} {\bf 81}(3) 034312

\bibitem{CTP12}
Carlsson B~G, Toivanen J and Pastore A 2012 {\em Phys. Rev. C\/} {\bf 86}(1)
  014307

\bibitem{NIY07}
Nakatsukasa T, Inakura T and Yabana K 2007 {\em Phys. Rev. C\/} {\bf 76} 024318

\bibitem{INY09}
Inakura T, Nakatsukasa T and Yabana K 2009 {\em Phys. Rev. C\/} {\bf 80} 044301

\bibitem{INY11}
Inakura T, Nakatsukasa T and Yabana K 2011 {\em Phys. Rev. C\/} {\bf 84}
  021302

\bibitem{AN11}
Avogadro P and Nakatsukasa T 2011 {\em Phys. Rev. C\/} {\bf 84} 014314

\bibitem{Sto11}
Stoitsov M, Kortelainen M, Nakatsukasa T, Losa C and Nazarewicz W 2011 {\em
  Phys. Rev. C\/} {\bf 84} 041305

\bibitem{AN13}
Avogadro P and Nakatsukasa T 2013 {\em Phys. Rev. C\/} {\bf 87} 014331

\bibitem{LNZM13}
Liang H, Nakatsukasa T, Niu Z and Meng J 2013 {\em Phys. Rev. C\/} {\bf 87}
  054310

\bibitem{HKN13}
Hinohara N, Kortelainen M and Nazarewicz W 2013 {\em Phys. Rev. C\/} {\bf
  87} 064309

\end{thebibliography}

\end{document}